# Molecular dynamics simulation of rapid solidification of Aluminum under pressure

A. Sarkar[*], P. Barat and P. Mukherjee

Variable Energy Cyclotron Centre, 1/AF Bidhan Nagar, Kolkata 700 064, India

[*]Corresponding author Email: *apu@veccal.ernet.in*

*Abstract*

*Molecular dynamics simulation study based on the EAM potential is carried out to investigate the effect of pressure on the rapid solidification of Aluminum. The radial distribution function is used to characterize the structure of the Al solidified under different pressures. It is indicated that a high pressure leads to strong crystallization tendency during cooling.*

## INTRODUCTION

A metallic liquid can form either a non-equilibrium phase (a metallic glass or quasicrystal) which is obtained by rapid quenching so that the nucleation of equilibrium phase is suppressed, or alternatively, an equilibrium phase, the crystal if the quenching process is slow. The pressure is also a critical factor affecting the resulting structure of a liquid-solid transition. Over the past few years, many works have been carried out to investigate the cooling rate dependence of the solidification structure. But few works can be found on the pressure induced structural changes during solidification. It is difficult to study the pressure effect on glass transition experimentally because the undercooled liquid has a poor thermal stability against crystallization. Computer simulation studies provide important complementarities to experimental studies, so that the understanding of atomic configuration is achieved in details. The molecular dynamics (MD) simulation has become a popular tool for investigation of liquid structure theoretically. In this paper we carry out MD simulation to investigate a cooling process of Aluminum (Al) under normal and high pressure.

**POTENTIAL ENERGY FUNCTION**

MD is a direct simulation technique at the atomic level. Almost all the physical properties of the material may be determined using molecular dynamics. It requires generally an inter-atomic potential. The atomic trajectories are calculated from the equation of motion by supposing a functional form of the inter-atomic potential. Recent years have witnessed considerable progress in the development of empirical and semi-empirical many-body potentials for MD simulations. The embedded-atom method (EAM) of Daw and Baskes [1,2] (and further developed by Johnson [3,4]), N-body potentials proposed by Finnis and Sinclair [5] and the tight-binding model of many body potentials given by Rosato et. al. [6] consist of three main aspects of the present empirical and semi-empirical many-body potentials. The EAM potential has proven to reproduce very well the basic structural and dynamical properties of solids, surfaces, defects, liquids etc.

The embedded-atom method (EAM) is used considering that it has been validated [7]. In this work, we have simulated the rapid solidification of Al under pressure using a realistic embedded-atom potential. In the framework of the EAM method, the total energy of an N-atoms system takes the form:

$$E_{tot} = \sum_i F_i(\rho_i) + \frac{1}{2}\sum_{i \neq j} \phi_{ij}(r_{ij}) \qquad (1)$$

where $\phi_{ij}(r_{ij})$ is a two-body central potential between atom $i$ and $j$ with the separation distance $r_{ij}$, $F_i(\rho_i)$ is the embedding energy, $\rho_i$ is the electronic density at atom $i$ due to the surrounding atoms:

$$\rho_i = \sum_{j \neq i} f_j(r_{ij}) \qquad (2)$$

where $f_j(r_{ij})$ is the contribution to the electronic density at atom $i$ due to atom $j$ at a distance $r_{ij}$ from the atom $i$. This function for the given element is derived from quantum-chemical calculations or from empirical information. Functions $\phi_{ij}(r_{ij})$ and $F(\rho)$ are selected such as to obtain agreement of calculated values with experimental data for a crystal (lattice parameter, elastic constant etc.). The following functional form of $F(\rho)$ is universally used for all metals

$$F(\rho) = -c\rho^{1/2} \qquad (3)$$

It allows one to interpret the $f(r_{ij})$ as an overlap integral of wave functions of atoms $i$ and $j$.

**MOLECULAR DYNAMICS SIMULATION**

Based on the constant temperature, constant pressure method (NPT-MD), the simulations are performed on a 5×5×5 cubic box subjected with periodic boundary conditions for a system of 500 atoms using the MD simulation package SAGE MD [8]. The typical time step is of the order of few femtoseconds. The equations controlling the motion of atoms are numerically solved using the Verlet algorithm [9]. The simulation is started at 1500 K, which is much higher than the melting temperature (933.5 K) of Al. We first run 10000 time steps at this temperature under the pressure 0, 0.5, 1, 2 GPa, respectively, to keep the system in a equilibrium state before cooling. The system is then cooled to the room temperature (300 K) under the same pressure at which the system is cooled at a cooling rate of $3 \times 10^{16}$ K/S.

The structural properties of the system are determined by using the radial distribution function (RDF) calculated as

$$g(r) = \frac{V}{N^2} \left\langle \frac{\sum_i n_i(r)}{4\pi r^2 \Delta r} \right\rangle \qquad (4)$$

Here, $r$ is the radial distance, $n_i(r)$ is the coordination number of atom $i$ separated by $r$ within $\Delta r$ interval, and the bracket denotes the time average. V is the volume of the simulation box and N is the number of atoms in the system. $g(r)$ basically gives the probability of finding an atom in a distance ranging from $r$ and $r+\Delta r$.

**RESULTS AND DISCUSSIONS**

Now it is well established that, when a liquid metal is cooled down, either crystalline solid or amorphous solid is obtained depending upon the cooling rate. If the cooling rate is sufficiently high, homogeneous nucleation of crystalline phase can be completely avoided and thus the metastable amorphous structure is formed. This concept has been a major breakthrough in preparing bulk metallic glasses, which have tremendous practical applications. In this study we move one step forward and intend to investigate the effect of pressure on the rapid solidification of a metal (Al) from its liquid state. An Al system has been first heated to 1500 K and then rapid cooled (cooling rate of $3\times10^{16}$ K/S) to 300 K. The heating and cooling have been performed at different pressures. Fig. 1 shows a typical variation of the temperature of the system against time.

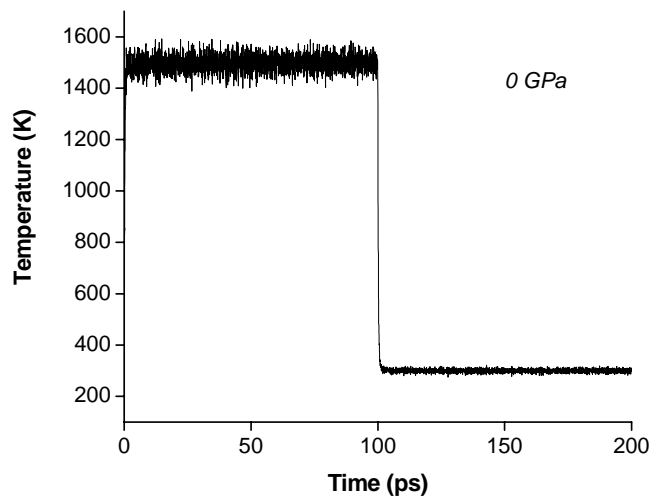

**Fig 1** Change of temperature during fast cooling.

It is seen that the system attained equilibrium temperatures 1500 K prior to cooling and 300 K after cooling. To investigate the structure of the liquid state and the quenched state we have calculated the RDF for both states. Here, it is worth to mention that RDF of a system can also be obtained experimentally from the diffraction studies. But, the experimental investigation of the effect of rapid cooling is a very difficult task. Fig 1 shows the RDF of the Al at 1500 K at different pressures.

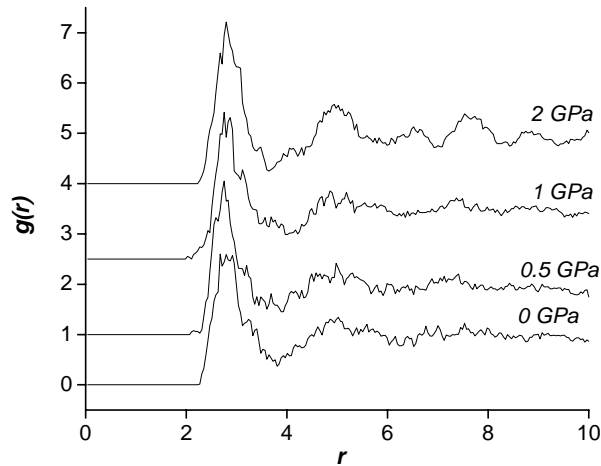

**Fig 2** The RDF of Al at 1500 K at various pressures

First we discuss the RDFs for pressures 0, 0.5 and 1 GPa. For these pressures the RDFs show typical short range order of a liquid state. The first peak of $g(r)$ is salient and the second one is smooth. On the contrary, the RDF at 2 GPa pressure shows little bit different behavior. Here the extent of ordering is more. Fig 3 shows the RDFs of the Al solidified under different pressures at a cooling rate of $3\times10^{16}$ K/S. The principal structure features like amorphous solid is clearly seen in the RDF at the cooled state (300 K) at pressures 0, 0.5 and 1 GPa. The splitting of the second peak of the $g(r)$ gives evidence for the appearance of an amorphous structure. It is seen that with increasing pressure the peaks of the RDF become sharper, indicating that the order degree of the system is strengthened and crystallization begins. Moreover the first peak in the RDF of the cooled state of Al formed under higher pressure shows an inward shift. This means that higher pressure decreases the distance between atoms. The RDF of the cooled state for 2 GPa pressure exhibits strong peaks indicating the appearance of crystalline phase at this pressure.

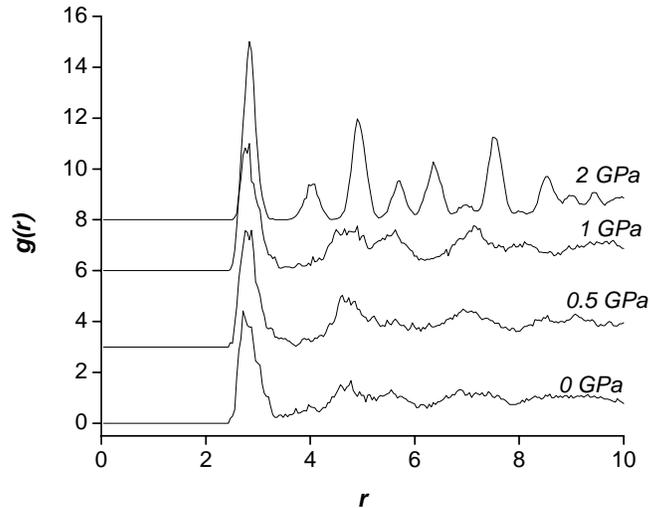

**Fig 3** The RDF of Al solidified (300 K) under different pressure.

The previous MD simulation studies of the pressure effect on the rapid solidification were done on Cu [10] and Au [11]. In those cases the crystallization was found to appear at much higher pressure. But in the case of Al the crystallization appeared at 2 GPa. This may be attributed to the low bulk modulus of Al (76 GPa) compared to Cu (140 GPa) and Au (220 GPa).

**CONCLUSION**

The structural evolution of bulk aluminum system during rapid cooling under same cooling condition at different pressures is investigated by the molecular dynamics simulation. Radial distribution function is adopted to explore the structural changes. The study reveals the fact that the induced pressure during solidification takes the Al in crystalline form even in very high cooling rate.

**REFERENCES**

[1] M. S. Daw, M. I. Baskes, Phys. Rev. Lett. **50** (1983) 1385.

[2] M. S. Daw, M. I. Baskes, Phys. Rev. B. **29** (1984) 6443.

[3] R. A. Johnson, Phys. Rev. B. **37** (1988) 3924.

[4] R. A. Johnson, Phys. Rev. B. **39** (1989) 12554.

[5] M. W. Finnis, J. E. Sinclair, Philos. Mag. A 50 (1984) 45.

[6] V. Rosato, M. Guillope, B. Legrand, Philos. Mag. A 59 (1989) 321.

[7] M.I. Baskes, Phys. Rev B, **46** (1992) 2727

[8] A.A. Selezenev et. al., Comp. Mater. Sci. **28** (2003) 107.

[9] L. Verlet, Phys. Rev. B **159** (1967) 98.

[10 ]J. Liu, J.Z. Zhao, Z.Q. Hu, Comp. Mater. Sci. **37** (2006) 234.

[11] Y.N. Zhang et. al., Phys. Lett. A **320** (2004) 452.